%% file: main.tex
\documentclass[5p, times, final]{elsarticle}
\usepackage{times}
\usepackage{epsfig}
\usepackage{graphicx}
\usepackage{amsmath}
\usepackage{amssymb}
\usepackage{color}
\usepackage{siunitx}
\usepackage{subcaption}
\usepackage{todonotes}
\usepackage{multirow}
\usepackage{graphicx}

\usepackage[pagebackref=true,breaklinks=true,colorlinks,bookmarks=false]{hyperref}

\usepackage{lineno,hyperref}
\modulolinenumbers[5]
\DeclareSIUnit\pixel{pixel}

\journal{Neurocomputing}

\begin{document}

\begin{frontmatter}

\title{Deep Learning-based Aerial Image Segmentation with Open Data for Disaster Impact Assessment }
\author{Ananya Gupta, Simon Watson, Hujun Yin}
\address{Department of Electrical and Electronic Engineering, The University of Manchester, Manchester, United Kingdom}




\begin{abstract}
Satellite images are an extremely valuable resource in the aftermath of natural disasters such as hurricanes and tsunamis where they can be used for risk assessment and disaster management. In order to provide timely and actionable information for disaster response, in this paper a framework utilising segmentation neural networks is proposed to identify impacted areas and accessible roads in post-disaster scenarios. The effectiveness of pretraining with ImageNet on the task of aerial image segmentation has been analysed and performances of popular segmentation models compared. Experimental results show that pretraining on ImageNet usually improves the segmentation performance for a number of models. Open data available from OpenStreetMap (OSM) is used for training, forgoing the need for time-consuming manual annotation. The method also makes use of graph theory to update road network data available from OSM and to detect the changes caused by a natural disaster. Extensive experiments on data from the 2018 tsunami that struck Palu, Indonesia show the effectiveness of the proposed framework. ENetSeparable, with 30\% fewer parameters compared to ENet, achieved comparable segmentation results to that of the state-of-the-art networks.  


\end{abstract}


\end{frontmatter}


\section{Introduction}

Satellite imagery is an extremely important resource for disaster management and response. Following a major natural disaster such as an earthquake or a tsunami, authorised users from national civil protection, rescue or security organisations can activate the International Charter: Space and Major Disasters~\cite{Boccardo2015}. The Charter is a worldwide collaboration amongst space agencies and space systems operators that provide satellite imagery for disaster monitoring. This imagery can then be used to identify damaged areas that need the most support and also routes that are still accessible for evacuation and emergency responses.

Such image analysis is typically done manually with support from volunteer initiatives such as the Humanitarian OpenStreetMap (OSM) team. They organise mapathons with volunteers from around the world to manually annotate high resolution satellite images. Inevitably, this process can be slow and error-prone due to the inexperience of many volunteers~\cite{Poiani2016}. Time is extremely critical in post-disaster situations for prompt relief efforts. Timely and accurate road maps are also extremely important for navigation in post-disaster scenarios. Pre-existing maps can be rendered inaccurate due to possible route blockages, water-logging, landslides and structural damages.

\begin{figure}[]
\centering
\begin{subfigure}[]{0.23\textwidth}
\centering
  \includegraphics[width=\textwidth]{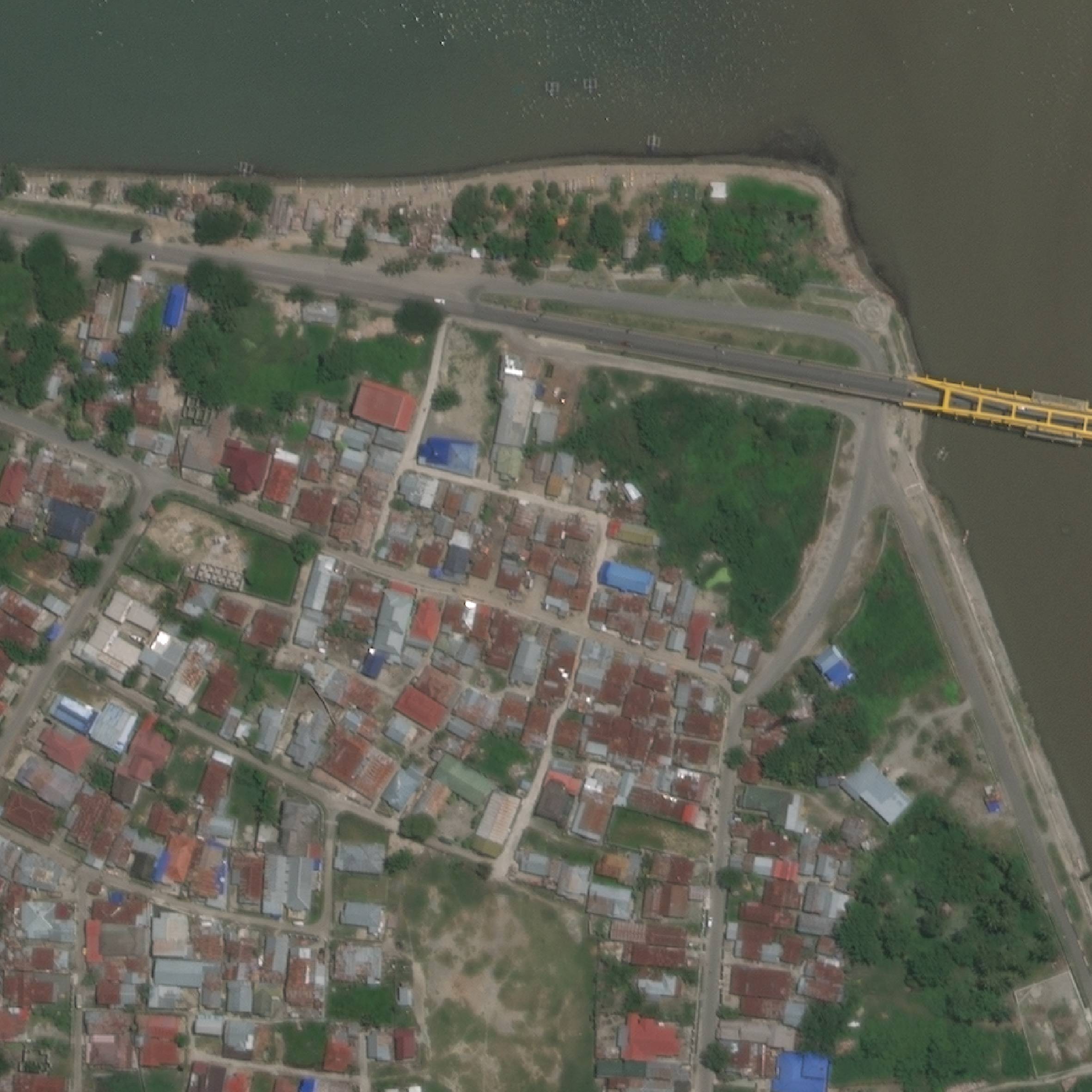}
\end{subfigure}
\begin{subfigure}[]{0.23\textwidth}
  \centering
  \includegraphics[width=\textwidth]{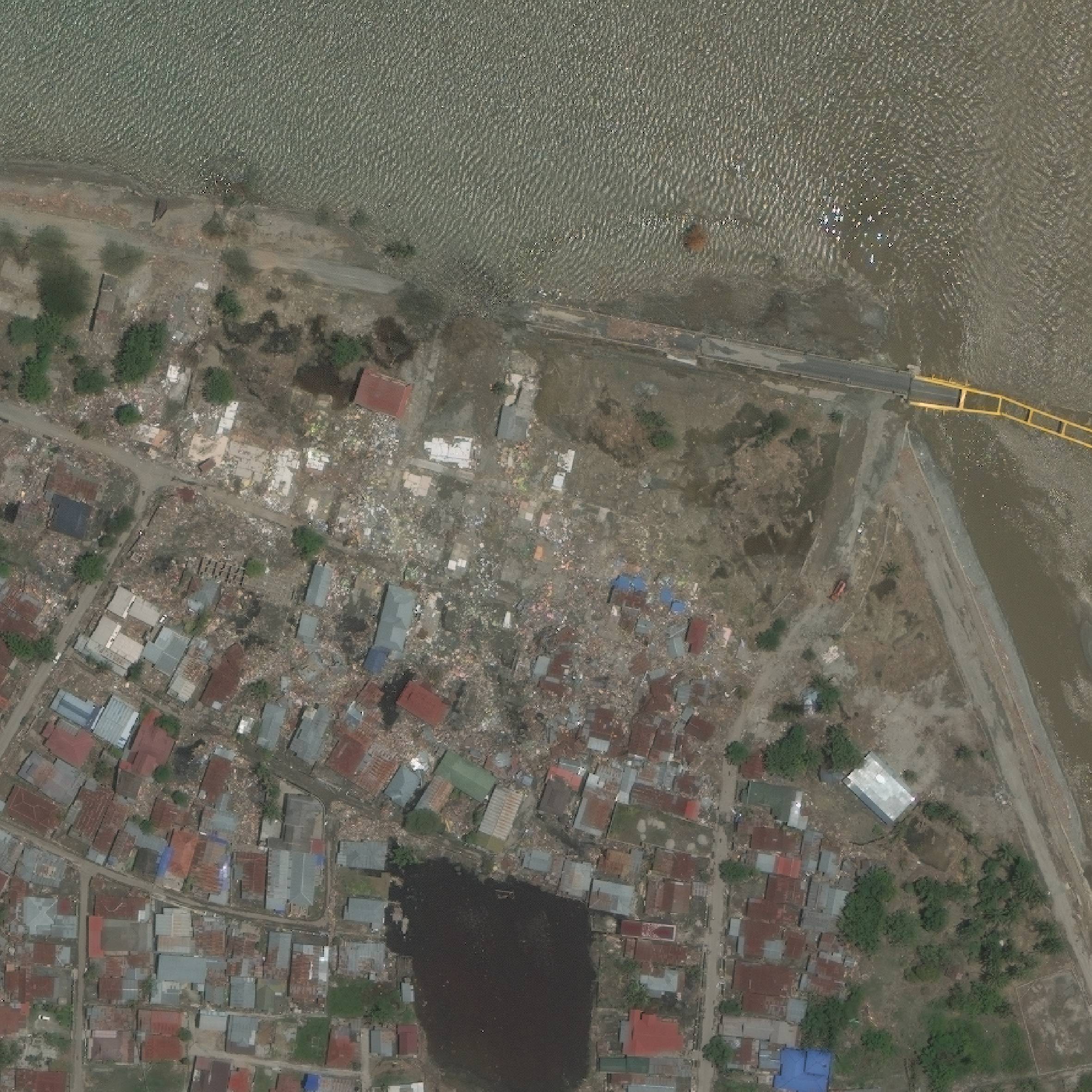}
\end{subfigure}
\caption{Extracted images from satellite imagery of Palu, Indonesia showing the devastation due to the tsunami and earthquake in September, 2018~\cite{DIGI}. \textit{Left}: Before the tsunami. \textit{Right}: The day after the tsunami.}
\label{fig:dev}
\end{figure}

For instance, on September 28, 2018, a 7.5 magnitude earthquake with an epicenter in Central Sulawesi struck Indonesia and led to a tsunami in the province capital Palu, which washed away a lot of the coastal infrastructure. This was the deadliest earthquake worldwide in 2018, with over 4,000 fatalities and damages to over 60,000 buildings. Following the event, a number of rapid mapping efforts were initiated by the government and volunteer initiatives for damage assessment. However, these efforts took days to complete due to many manual processes~\cite{Adriano2019}.

Deep learning (DL) based techniques such as convolutional neural networks (CNN) are becoming increasingly pervasive as a means of automating the knowledge discovery process in many fields including remote sensing~\cite{Zhu2017}. These techniques are used in conjunction with Earth Observation data for applications such land-use classification, change detection, object detection and disaster analysis. However, DL models are data driven and typically require a large amount of manually annotated data for training~\cite{Adriano2019}. This data annotation is a slow process and hence these methods cannot be directly used for rapid disaster analysis.

Some recent work has explored the use of data from OSM~\cite{Contributors2017} for training ML models in the absence of high quality manually labelled training data~\cite{Kaiser2017}. OSM data can be assumed to be weakly labelled training data due to issues such as mis-registration and out-of-date labels. The work in \cite{Kaiser2017} showed that a large enough training dataset helped alleviate the issues of using a weakly labelled dataset. 

A framework to detect damaged roads from satellite imagery and register them to OSM was recently proposed in \cite{Gupta2019b}. It was trained on publicly available OSM data, forgoing the requirement for expensive manually annotated data. It provided a method inspired by graph theory to automate the process of updating the OSM database by combining the changes detected using their framework with OSM road data. This paper builds on \cite{Gupta2019b} and shows how it can be extended to multiple semantic classes. A systematic analysis of the performance of different neural networks for the task of aerial image segmentation is performed. The effect of pretraining the neural networks on a large image dataset is also analysed. Two variants of popular neural networks are also proposed: the first, ENetSeparable, focuses on efficiency and the second, UNetUpsample, on accuracy. Finally, it is shown that the proposed framework from \cite{Gupta2019b} can be seen as being architecture agnostic since it helps reduce the difference in segmentation performance from the various segmentation networks.





\section{Related Work}

\subsection{Aerial Image Segmentation}

Recent successes of deep learning models in image classification and big data analysis have promoted much increased use of such models in remote sensing for tasks such as land cover classification and change detection~\cite{Zhu2017}. Readers are referred to \cite{Zhang2016b} and \cite{Zhu2017} for an extensive background and review on the use of deep learning for remote sensing tasks. Common tasks in this field are extraction of road networks~\cite{Bastani2018,Mattyus2017,Sun2018} and building footprints~\cite{VanEtten2018a} using semantic segmentation networks, popularised by large-scale competitions such as DeepGlobe~\cite{Demir2018} and SpaceNet~\cite{VanEtten2018}.

Most popular semantic segmentation architectures are structured as encoder-decoder models popularised by UNet~\cite{Ronneberger2015}. The encoder consists of a number of blocks where each block takes an input image or feature map and produces a set of downsampled feature maps which progressively identify higher level features. The decoder network mirrors the encoder network and progressively upsamples the output from the encoder network. Individual decoder blocks are connected to the corresponding encoder blocks with skip links to help recover the fine-grained details lost in the downsampling. The upsampling is typically done using transposed convolutions with learnable weights. 

A study on using OSM data for learning aerial image segmentation showed that using a large amount of weakly labelled data for training helped achieve reasonable performance without the need for large well-labelled datasets~\cite{Kaiser2017}. Alternative schemes to train models for aerial image segmentation employ  self-supervision~\cite{Singh2018} and supervised pretraining on the ImageNet~\cite{Audebert2016}. 

\subsection{Disaster Analysis}

Remote sensing is being increasingly used for disaster response management due to the increasing availability of remote sensing data, which can be acquired relatively quickly~\cite{Li2016}. The main datatypes used in such cases are synthetic aperture radar (SAR) and high resolution optical images. SAR is extremely useful for dealing with low-light conditions and for areas with cloud cover. It is especially useful in finding flooded areas~\cite{Schumann2007} and identifying ground displacements after earthquakes~\cite{Pathier2006}. However, it cannot be used in urban areas with the same effectiveness due to radar backscattering to the sensors caused by tall objects such as buildings~\cite{Boccardo2015}.

High resolution optical imagery is typically used for visual interpretation in the case of events such as hurricanes, cyclones and tsunamis, which leave visible damages to an area. Recent work in automating this process has focused on assessing the damage to buildings in disaster-struck areas. A combination of pre- and post-tsunami satellite images has been used to assess whether a building was washed away by a flood with the implementation of a CNN~\cite{Fujita2017}. Similarly, an approach fusing multi-resolution, multi-sensor and multi-temporal imagery in a CNN was used to segment flooded buildings~\cite{Rudner2019}. However, these approaches require manually labelled post-disaster data for training their networks, which is time-consuming and expensive to obtain.

Automated road extraction from satellite imagery is an area of interest since a number of location and navigation services require up-to-date road maps~\cite{Miller2014}. A number of approaches using segmentation methods have been proposed to extract road networks. These methods typically depend on heuristics-based techniques in post processing to fix incorrect gaps from the segmentation networks~\cite{Mattyus2017}. However, these methods, while valid in typical road extraction scenarios, are not suitable in post-disaster scenarios because gaps in the segmentation masks could be caused by the effects of a disaster and are of extreme importance. 

There is some existing research for road extraction in post-disaster scenarios. Vehicle trajectories have been used for identifying obstacles such as standing waters and fallen trees~\cite{Chen2018}. Road centerline extraction from post-disaster imagery has used OSM vector data for generating seed points and creating a more accurate road map following an earthquake, which can cause registration errors~\cite{Liu2019}. This method only corrects the registration errors but does not deal with the problem of destroyed roads. A crowd-sourced pedestrian map builder has also been developed~\cite{Bhattacharjee2019} but it requires people walking around in potentially destroyed areas and is not scalable.

Segmentation networks have also been used for detecting changes caused by disasters~\cite{Doshi2018}. These methods use the difference between outputs of pre-disaster and post-disaster imagery to obtain a measure on areas that have been damaged the most. In contrast, the current work extends the previous work by identifying the changes to the road networks at a fine-grained level. Furthermore, the proposed framework also allows for an update to OSM to achieve more realistic road network maps for the area under consideration.

\section{Methodology}

The proposed disaster impact assessment is based on finding the difference in roads and buildings between satellite imagery from before and after a disaster. This is done by using a semantic segmentation network trained on pre-disaster aerial imagery for identifying these objects in the before and after imagery. The difference in the predicted road masks is further used to update data from OSM for finding accessible routes in the post-disaster scenario.

\subsection{Segmentation Models} 
\label{sec:segmentation}


The models used in this study are modified versions of the UNet and LinkNet~\cite{Chaurasia2018}. The modifications were inspired by the TernausNet~\cite{Iglovikov2018}, which showed that replacing the UNet encoder with a pretrained VGG11 encoder improved segmentation results. 

Here a systematic study was carried out to compare the effectiveness of different encoder backbones. In the tested models, the encoder backbone was replaced by the convolutional layers from VGG~\cite{Simonyan2015} and ResNet~\cite{He2015} for the UNet. The original LinkNet model, with ResNet18 as its encoder, and another one with a ResNet34 backend were also tested.

A slight modification of the UNet was also studied where the transposed convolutions in the decoders were replaced with the nearest neighbour upsampling to deal with possible checkerboard artifacts~\cite{Odena2016}. This modified version has been called UNetUp in the remainder of the text.



Another model tested in this study is the ENet~\cite{Paszke2016}. It is an encoder-decoder model optimised for efficiency in terms of latency and parameters, with an encoder inspired by ResNet and a small decoder. It uses early downsampling with a relatively low number of feature maps to reduce the number of operations required. It also decomposes \(n\times n\) convolutions into smaller convolutions of \(n \times 1\) and \(1 \times n\)\cite{Jin2014}, allowing for large speedups. 

Inspired by Xception-Net~\cite{Chollet2016}, a modified version of ENet, called ENetSeparable, is proposed. In this model, all convolutional filters are replaced by depthwise separable convolutions, a modification that reduces the number of parameters by 30\%.

The loss function is a weighted cross entropy loss with an additional soft Jaccard constraint and is given as follows:

\begin{equation}
\label{eq:loss}
    L = (1-\alpha) \frac{1}{I}\sum_i^I \left(-w_k log \frac{e^{o_{ik}}}{\sum_c^C e^{o_{ic}}}\right) - \alpha \sum_c^C log \frac{\sum_{i}e^{o_{ic}}*t_{ic}}{\sum_{i}e^{o_{ic}} + t_{ic} - e^{o_{ic}}*t_{ic}}
\end{equation}{}
where

\begin{equation*}
\label{eq:weights}
    w_k = \frac{\sum_{c=1}^{C}S_c}{C\times S_k}
\end{equation*}{}

Subscript \(k\) denotes the target class, \(i\) indexes over all pixels where the total number of pixels is given by \(I\) and the total number of classes is given by \(C\). The output for class \(c\) at pixel \(i\) is given by \(o_{ic}\) and \(t\) is a one-hot encoded target vector. \(\alpha\) is the weighting for the Jaccard loss. The weight for class \(k\) is given by \(w_k\) , and \(S_k\) denotes the number of samples in the training set for target class \(k\) while \(S_c\) the number of samples for class \(c\).

.

\subsection{Disaster Impact Assessment with Change Detection}
\label{sec:mask_diff}
The segmentation network is used to identify buildings and roads in pre-disaster and post-disaster aerial imagery. Due to shadows and occlusions, the segmentation output can have a number of incorrect gaps. The building and road segmentation masks are dilated with a small kernel (e.g. \(5\times5\)) for several iterations (6 in our experiments) to overcome some of these gaps. The resulting masks can be used to distinguish the infrastructure that was destroyed due to the disaster as follows:

 \begin{equation}
 \label{eq:mask_diff}
     M_{diff_{p}} = 
     \left\{\begin{matrix}
      1 & \text{if} \quad\! M_{pre_{p}}\in {1,2} \quad\! \textrm{and} \quad\! M_{post_{p}} = 0 \\
      0 & \text{otherwise}
\end{matrix}\right.
\end{equation}

\begin{figure*}[]
    \centering
    \includegraphics[width=0.9\textwidth]{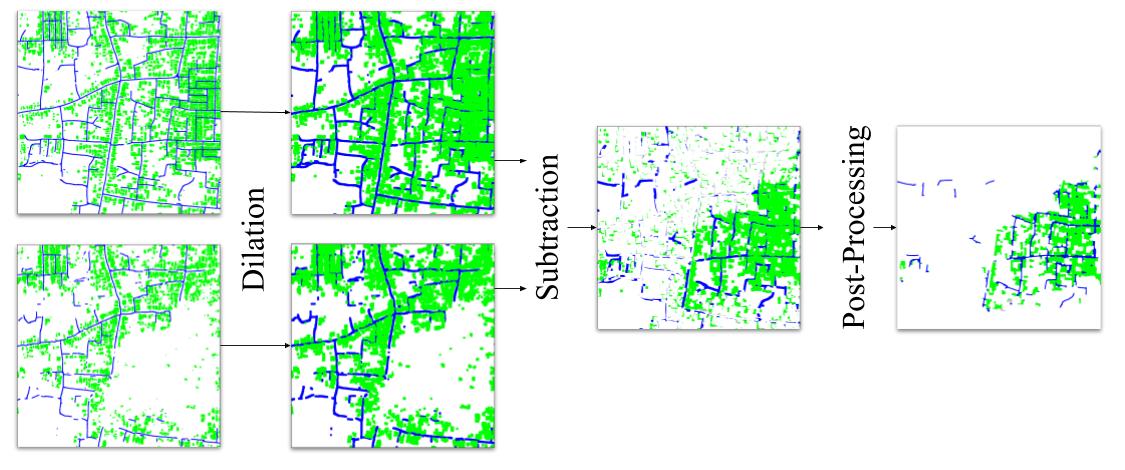}
    \caption{Pipeline for change detection in pre-disaster and post-disaster segmentation masks. Roads shown in blue and buildings shown in green.}
    \label{fig:mask_diff}
\end{figure*}{}

Subscript \(p\) indexes over all the pixels in each image and \(M_{diff}\) is the disaster difference mask. \(M_{pre}\) and \(M_{post}\) are the segmentation masks from pre and post-disaster imagery respectively. The inferred label is one of {0,1,2} referring to the background, building or road class. This function computes true for any pixel that was identified as a road or building in the pre-disaster image but as background in the post-disaster image since that can be assumed to be damaged due to the disaster. 

Due to small mis-registration issues and non-ideal segmentation outputs, the segmentation masks from the pre-disaster and post-disaster images do not completely overlap. Hence, small blobs in the difference mask can be assumed to be noise or artifacts caused by the registration error. Morphological erosion and opening are used to remove all such noise and the final mask obtained represents the damaged infrastructure due to the disaster. The intermediate steps of this process are shown in Fig. \ref{fig:mask_diff}.

\subsection{Generating Road Graphs}
\label{sec:road_graph} 

The output segmentation mask is converted to a road network graph motivated by graph theory to obtain a map suitable for route computation. Firstly, all pixels marked as road are extracted to form a road mask. The road mask is dilated to deal with small gaps in the segmentation output since these can cause large errors in the network graph. Morphological thinning is performed on the obtained mask to get a single pixel thick road skeleton. The road skeleton is traversed to find all nodes where each node is any positive pixel with three or more positive pixel neighbours. All pixels between two nodes are marked as part of an edge. 

Since the edges approximated with this method are fairly crooked and small road segments can be assumed to be straight, the edges are simplified to piece-wise linear segments using the Ramer-Douglas-Pecker algorithm~\cite{DOUGLAS1973}.

\subsection{Registering changes to OSM}
\label{sec:graph_comp}

The road network generated from post-disaster imagery using the methodology described above could be used for routing in most scenarios. However, non-ideal segmentation masks can cause long detours when creating the road network graph. Hence, it is proposed to further use data from OSM as the best estimate of the world prior to a disaster and register the changes caused by such an event with the OSM road graph to obtain an updated map of the affected region. The change graph can be obtained from the difference mask generated in Section \ref{sec:mask_diff}. Note that the OSM data is not completely accurate~\cite{Mattyus2017}, but based on empirical observations, using it provides more robust results.

There are a number of methods in graph theory for measuring graph similarity. However, these methods compare logical topology of graphs by looking for common nodes. In the case of road networks, the physical topology is extremely important and the graph comparison problem becomes non-trivial. In such cases, corresponding nodes in the two graphs may not spatially coincide due to image offsets and errors in the segmentation masks making the pre-existing methods of graph comparison unfeasible.

\begin{figure}[]
    \centering
    \includegraphics[width=0.5\textwidth]{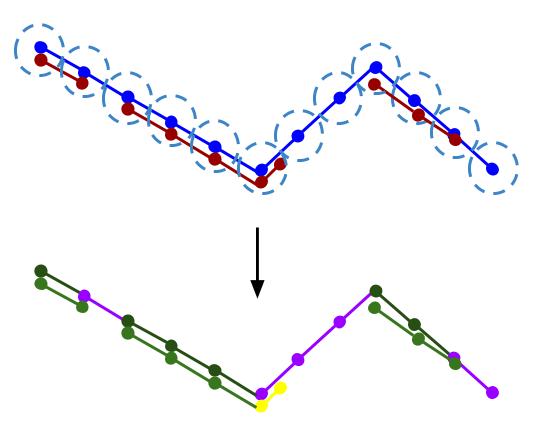}
    \caption{Graph Comparison. \textit{Top}: \(G'_a\) in blue and \(G'_b\) in red with dashed circles of radius l/2 drawn around the nodes of \(G'_a\). \textit{Bottom}: Common sub-segments shown in green, non-corresponding sub-segments from \(G'_a\) shown in purple and those from \(G'_b\) shown in yellow. }
    \label{fig:graph_comp}
\end{figure}{}

In order to compare the topological graphs, each edge of the graphs \(G_a\) and \(G_b\) is sliced into smaller sub-segments of length \(l\) to obtain simplified graphs \(G'_a\) and \(G'_b\). Corresponding sub-segments in the two graphs can be found using Eq. \ref{eq:graph_comp}~\cite{Gupta2019b}, where two sub-segments are assumed to be corresponding if both vertices of one sub-segment are within a certain distance of the other sub-segment. A visual representation of this can be seen in Figure \ref{fig:graph_comp}.

\begin{equation}
\label{eq:graph_comp}
\begin{split}
    \forall e_a, e_b; e_a \in G'_a,  e_b \in G'_b \\
    e_a = \{v_{a1}, v_{a2}\}; e_b = \{v_{b1}, v_{b2}\} \\
    e_a = e_b, \quad \! \textrm{iff} \quad \! |a1 - b1| < l/2 \quad\! \textrm{and} \quad\! |a2 - b2| < l/2
\end{split}
\end{equation}

In Eq. \ref{eq:graph_comp}, \(e_a\) and \(e_b\) are the sub-segments in graphs \( G'_a\) and \( G'_b\) and are defined in terms of their two vertices, \( v_{a1}\) and \(v_{a2}\), and \( v_{b1}\) and \(v_{b2}\), respectively. The euclidean distance between two vertices is given by \(|a1 - b1|\) where \(a1\) and \(b1\) represent the coordinates of the first vertices of \( v_{a1}\) and \( v_{b1}\), respectively.

\section{Experimental Setup}

\subsection{Neural Network Structures}

The structures of the encoders and decoders are summarised in Table \ref{tab:encoders}  and Table \ref{tab:decoders}, respectively. In these tables conv\(x\)-\(y\) implies a convolutional layer with a kernel size of \(x\) and \(y\) filters with the \(nc\) in the final layer meaning the number of output classes. Similarly, convTran\(x\)-\(y\) is a transposed convolution layer with a kernel size of \(x\) and \(y\) filters. The individual decoder block structure for the different models is given in Table \ref{tab:decoder_block}. All convolutional layers use the ReLU\cite{Nair2010} activation and the encoders include pooling and batch-norm layers as proposed by the original authors. Note that UNet-style architectures concatenate the encoder feature map and the decoder feature map, while LinkNet architectures add the feature maps instead of concatenating them to make the network more efficient. 

\begin{table*}[]
\caption{Encoder Structures}
\label{tab:encoders}
\centering
\begin{tabular}{|l|l|l|l|l|l|l|}
\hline
\textbf{Block} & \textbf{VGG11} & \textbf{VGG16} & \multicolumn{2}{l|}{\textbf{ResNet18}} & \multicolumn{2}{l|}{\textbf{ResNet34}} \\ \hline

\multirow{2}{*}{\textbf{enc1}} & \multirow{2}{*}{conv3-64} & conv3-64 & \multicolumn{2}{l|}{\multirow{2}{*}{conv7-64}} & \multicolumn{2}{l|}{\multirow{2}{*}{conv7-64}} \\ \cline{3-3}
 &  & conv3-64 & \multicolumn{2}{l|}{} & \multicolumn{2}{l|}{} \\ \hline
 
\multirow{2}{*}{\textbf{enc2}} & \multirow{2}{*}{conv3-128} & conv3-128 & conv3-64 & \multirow{2}{*}{x2} & conv3-64 &  \multirow{2}{*}{x3} \\ \cline{3-4} \cline{6-6} 
 &  & conv3-128 & conv3-64 &  & conv3-64 &  \\ \hline
 
\multirow{3}{*}{\textbf{enc3}} & conv3-256 & conv3-256 & conv3-128 & \multirow{3}{*}{x2} & conv3-128 & \multirow{3}{*}{x4} \\ \cline{2-4} \cline{6-6}
 & \multirow{2}{*}{conv3-256} & conv3-256 & \multirow{2}{*}{conv3-128} &  & \multirow{2}{*}{conv3-128} &  \\ \cline{3-3}
 &  & conv3-256 &  &  &  &  \\ \hline
\multirow{3}{*}{\textbf{enc4}} & conv3-512 & conv3-512 & conv3-256 & \multirow{3}{*}{x2} & conv3-256 & \multirow{3}{*}{x6} \\ \cline{2-4} \cline{6-6}
 & \multirow{2}{*}{conv3-512} & conv3-512 & \multirow{2}{*}{conv3-256} &  & \multirow{2}{*}{conv3-256} &  \\ \cline{3-3}
 &  & conv3-512 &  &  &  &  \\ \hline
\multirow{3}{*}{\textbf{enc4}} & conv3-512 & conv3-512 & conv3-512 & \multirow{3}{*}{x2} & conv3-512 & \multirow{3}{*}{x3} \\ \cline{2-4} \cline{6-6}
 & \multirow{2}{*}{conv3-512} & conv3-512 & \multirow{2}{*}{conv3-512} &  & \multirow{2}{*}{conv3-512} &  \\ \cline{3-3}
 &  & conv3-512 &  &  &  &  \\ \hline
\end{tabular}
\end{table*}

\begin{table*}[]
\caption{Decoder Structures}
\label{tab:decoders}
\centering
\begin{tabular}{|l|l|l|l|}
\hline
\textbf{Block} & \textbf{UNet} & \textbf{UNetUp} & \textbf{LinkNet} \\ \hline
\textbf{center} & dec\_unet(512,256) & dec\_unet\_up(512,256) & dec\_link(512,256) \\ \hline
\textbf{dec5} & dec\_unet(512,256) & dec\_unet\_up(512,256) & dec\_link(256,128) \\ \hline
\textbf{dec4} & dec\_unet(256,128) & dec\_unet\_up(256,128) & dec\_link(128,64) \\ \hline
\textbf{dec3} & dec\_unet(128,64) & dec\_unet\_up(128,64) & dec\_link(64,64) \\ \hline
\textbf{dec2} & dec\_unet(64,32) & dec\_unet\_up(64,32) & convTran3-32 \\ \hline
\textbf{dec1} & conv3-32 & conv3-32 & conv3-32 \\ \hline
\textbf{final} & conv3-nc & conv3-nc & conv3-nc \\ \hline
\end{tabular}
\end{table*}

\begin{table}[]
\caption{Decoder Block Structure}
\label{tab:decoder_block}
\centering
\begin{tabular}{|l|l|}
\hline
\textbf{Block} & \textbf{Layers} \\ \hline
\multirow{2}{*}{dec\_unet(a,b)} & conv3-a \\ \cline{2-2} 
 & convTran4-b \\ \hline
\multirow{3}{*}{dec\_unet\_up(a,b)} & upsample \\ \cline{2-2} 
 & conv3-a \\ \cline{2-2} 
 & conv3-b \\ \hline
\multirow{3}{*}{dec\_link(a,b)} & conv3-a/4 \\ \cline{2-2} 
 & convTran4-a/4 \\ \cline{2-2} 
 & conv3-b \\ \hline
\end{tabular}
\end{table}

\subsection{Datasets}

DigitalGlobe's Open Data Program\footnote{https://www.digitalglobe.com/ecosystem/open-data} provides high resolution satellite imagery in the wake of natural disasters to enable a timely response. This study uses the data from Palu, Indonesia, which was struck by an earthquake and tsunami on 28 September, 2018 and had visible damages to its coastlines and infrastructure. The pre-disaster imagery was from 7th April, 2018 and the post disaster imagery was from 1st October, 2018. The imagery had a ground sampling distance of approximately \SI{50}{\centi\metre\per\pixel}. 

\begin{figure}[]
    \centering
    \includegraphics[width=0.49\textwidth]{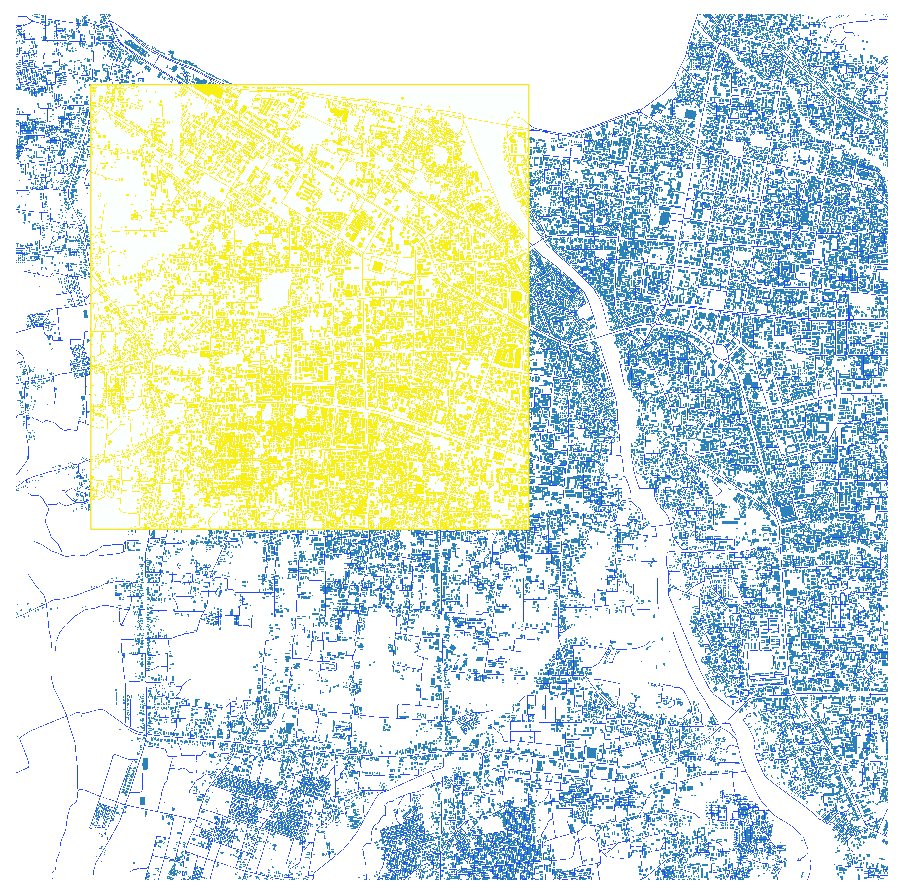}
    \caption{Dataset Extent: Training extent in blue and testing extent in yellow. The split was chosen such that most of the damaged area was part of the testing extent.}
    \label{fig:dataset_extent}
\end{figure}{}

An area of \SI{45}{\kilo\metre\squared} around Palu city was extracted for the experiments, with \SI{14}{\kilo\metre\squared} of the area with visible damage being set aside for testing. The remainder of the imagery was used for training and validation. The dataset split is visualised in Fig. \ref{fig:dataset_extent}, where the area in yellow was used for testing.

The labels for training the segmentation networks were downloaded from OSM\footnote{https://www.openstreetmap.org/}. All polylines marked as motorways, primary, secondary, tertiary, residential, service, trunk and their links were extracted as roads. The roads and buildings in OSM were provided as vectors and polygons, respectively. They were converted to a raster format to create a dataset suitable for training. All the lat-long coordinates were converted to pixel coordinates. The roads were rasterised with a buffer of \SI{2}{\metre} and the building polygons were rasterised as filled polygons. For the binary segmentation tasks, separate road and building mask images were generated where the target classes were labelled as 1. For the multiclass segmentation experiments, the buildings and roads were labelled as 1 and 2 respectively. The background pixels were always marked with 0. The test datasets were annotated manually.


Note that only pre-disaster data was used for training the neural networks and the segmentation based results. The post-disaster imagery was used purely for inference and for obtaining the post-disaster mapping results.

\subsection{Metrics}
\label{sec:metrics}

The Jaccard Index or Intersection over Union (IoU) is a typical per-pixel metric for evaluating segmentation results. It is given by Eq. \ref{eq:iou} and measures the overlap of predicted labels with the true labels. For the binary segmentation cases, the IoU for the target class is reported and for the multi-class case, the mean IoU (mIOU) over the target classes is also reported. The IoU for the background class is not included since the high number of background pixels would bias the results.

\begin{equation}
\label{eq:iou}
    IoU = \frac{TP}{TP+FP+FN}
\end{equation}{}

The IoU metric measures the segmentation performance but is not the most suitable metric for graphs, because a small gap in the segmentation mask may only cause a small error in the IoU metric but can lead to large detours if the resulting road network is used for navigation. As outlined in Section \ref{sec:graph_comp}, comparing two topological graphs is a non-trivial task and graph connectivity is as important as graph completeness. Herein, two metrics are used, the first to evaluate the completeness of the graph and the second to evaluate the connectivity of the generated graph. 

The first metric measures the similarity of the sub-segments described in Section \ref{sec:graph_comp} using the precision-recall metrics as follows:

\begin{equation}
\label{eq:pr}
\begin{split}
precision=\frac{TP}{TP+FP} \\
recall=\frac{TP}{TP+FN}  \\
F_{score} = 2\times \frac{p\times r}{p + r}
\end{split}
\end{equation}

The metric proposed in \cite{Wegner2015} has been reported for evaluating graph connectivity. This metric measures the similarity of graphs by comparing the shortest path length for a large set of random source-destination pairs between the actual graph and the predicted graph. If the extracted paths have a similar length, they can be assumed to be a match and are marked as 'Correct' in the results. If the path length in the predicted graph is smaller than the actual graph, the generated graph has incorrect connections and this is reported as 'Too Short'. Conversely, if there are incorrect gaps in the predicted graph, the paths are either 'Too Long' or there are no possible paths, giving 'No Connections'.

\subsection{Training Details}
The models were trained using the Adam optimiser~\cite{Kingma2015} with a learning rate of 10\(^{-4}\). A minibatch size of 5 images was used for all the UNet-based models and 32 images for the other models. The models were built in Pytorch~\cite{pytorch}. The VGG11, VGG16, ResNet18 and ResNet34 models provided by the Pytorch model zoo were used for initialising the encoder networks in the pretrained networks. He initialisation~\cite{He2015b} was used for all the other layers.
 
 The training images and their corresponding masks were cropped to 416\(\times\)416 pixels and were augmented with horizontal and vertical flipping. All images were zero-mean normalised. Only the pre-disaster images were used for training and the post-disaster images were used for inference.
 
 All models were trained for 600 epochs to enable a fair comparison between the different models. A validation set was used for preventing overfitting; the final model used for measuring the performance was set to be the one where the validation loss converged to just before starting diverging (i.e. overfitting on the training set).

\section{Results}

\subsection{Segmentation Results}

Our experiments tested the following scenarios:
\begin{itemize}
\item Effect of pretraining with ImageNet on aerial image segmentation models.
\item Efficiency vs. accuracy trade-off between different popular architectures.
\end{itemize}

\subsubsection{Effect of pretraining on aerial image segmentation}

The first set of experiments were conducted to analyse the effect of pretraining with ImageNet on the segmentation task. The purpose of these experiments was two-fold. Firstly, whether pretraining on a large classification dataset such as ImageNet improves the accuracy of an unrelated task where the image statistics are quite different (ground-based object images for classification vs. aerial images for segmentation). Secondly, assuming accuracy with pretraining is similar or even better than that without, whether pretraining improves the convergence speed.  

\begin{figure}[]
    \centering
    \includegraphics[width=0.49\textwidth]{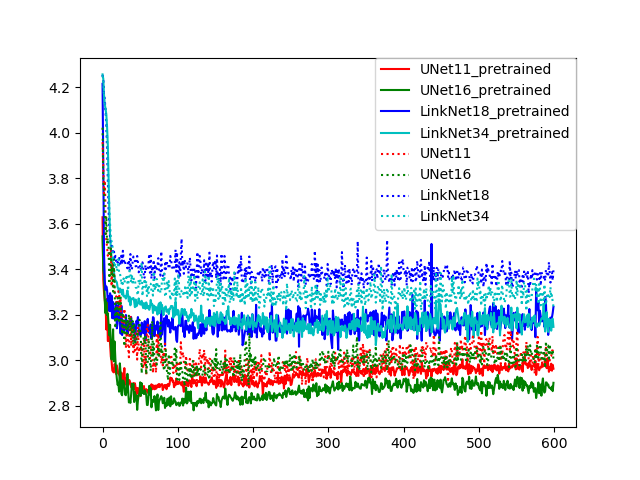}
    \caption{Buildings Validation Loss}
    \label{fig:building_pretrain_loss}
\end{figure}

\begin{figure}[]
    \centering
    \includegraphics[width=0.49\textwidth]{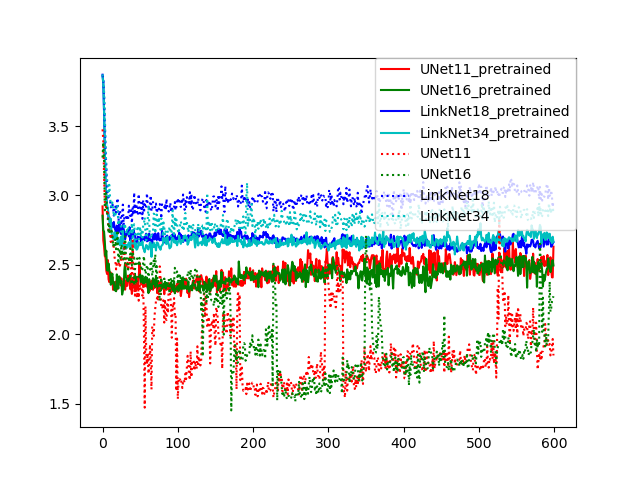}
    \caption{Roads Validation Loss}
    \label{fig:road_pretrain_loss}
\end{figure}

Four different network architectures were tested: UNet with VGG11 or VGG16 backend and LinkNet with ResNet18 or ResNet34 backend. These models were trained for segmentation of buildings and roads, while validation loss curves are shown in Fig. \ref{fig:building_pretrain_loss} and Fig. \ref{fig:road_pretrain_loss} respectively.  From the loss curves, it can be seen that pretrained networks generally converged quicker than their non-pretrained equivalents, requiring approximately 10 less epochs, regardless of the model used and the target class. The training curves for the VGG-based UNet models for the road segmentation task also show that these models did not converge very well when training from scratch, though their validation loss, which was used for preventing overfitting, was lower than their pretrained equivalent.

The results of these models on the test set, summarised in Table \ref{tab:pretraining_results}, show that in general the pretrained models had a higher IoU by a couple of points as compared to their non-pretrained equivalents; and this corresponds to the previous results in the literature~\cite{Audebert2016}. Note that this section focuses on identifying the differences between training models from scratch and using pretrained encoders. The results across different models are compared in the next section.

{}

{}

\begin{table}[]
\caption{Effects of pretraining. All results given as IoU}
\label{tab:pretraining_results}
\centering
\begin{tabular}{|l|l|l|l|}
\hline
\textbf{Model} & \textbf{Pretrained} & \textbf{Roads} & \textbf{Buildings} \\ \hline
UNet (VGG11) & No  &37.73 & 57.47   \\ \hline
UNet (VGG11) & Yes & 39.36 & 57.58   \\ \hline
UNet (VGG16)  & No   & 39.2 & 57.72 \\ \hline
UNet (VGG16)  & Yes  & 40.07  & 59.72    \\ \hline
LinkNet (ResNet18)  & No & 32.3 & 51.29 \\ \hline
LinkNet (ResNet18) & Yes & 35.08 & 57.08  \\ \hline
LinkNet (ResNet34) & No & 35.42 & 54.82  \\ \hline
LinkNet (ResNet34) & Yes & 37.2 & 57.15  \\ \hline
\end{tabular}
\end{table}


\subsubsection{Model capacity, design and accuracy}

 Visualisation of binary segmentation results for different models can be seen in Figure \ref{fig:building_seg} and Figure \ref{fig:road_seg} and their quantitative performance are reported in Table \ref{tab:seg_results}. 
Sizes of these different models are reported in Table \ref{tab:model_size}. From the results it can be seen that the proposed UNetUp with a VGG16 encoder outperformed all other models by a couple of points on each task. 

The building segmentation masks in Figure \ref{fig:building_seg} show that the ENet-based models led to fairly blob-like outputs without clear boundaries. The LinkNet models gave more distinct boundaries but the clearest results were with the UNet and UNetUp models. The road segmentation masks did not appear as distinctively different in terms of visual comparison, though the ENet based models seemed to miss the most segments in this case.

\begin{table*}[]
\caption{Segmentation Results (IoU) using pretrained models}
\label{tab:seg_results}
\centering
\begin{tabular}{|l|l|l|l|l|l|}
\hline
\multicolumn{1}{|c|}{\multirow{2}{*}{\textbf{Model}}} & \multicolumn{2}{c|}{\textbf{Binary}}& \multicolumn{3}{c|}{\textbf{Multiclass}}                                                                              \\ \cline{2-6} 
\multicolumn{1}{|c|}{}& \multicolumn{1}{c|}{\textbf{Roads}} & \multicolumn{1}{c|}{\textbf{Buildings}} & \multicolumn{1}{c|}{\textbf{Roads}} & \multicolumn{1}{c|}{\textbf{Buildings}} & \multicolumn{1}{c|}{\textbf{Average}} \\ \hline
UNet (VGG11)  & 39.36 & 57.58  &  35.91 & 51.92     & 43.92   \\ \hline
UNet (VGG16)   & 40.07  & 59.72    & 35.12   & 52.23    & 43.68  \\ \hline
UNet (ResNet18)   & 36.43 & 57.90 & 30.79 & 50.71  & 40.75\\ \hline
UNet (ResNet34)    & 37.56 & 58.29 & 34.18 & 51.08 &  42.63\\ \hline
LinkNet (ResNet18) & 35.08 & 57.08 & 29.25 & 50.59 & 39.92 \\ \hline
LinkNet (ResNet34) & 37.2 & 57.15 & 32.70 & 50.09 & 41.40 \\ \hline
ENet & 36.34 & 59.44 & - & - & - \\ \hline
ENetSeparable  & 37.44 & 59.58& - &  - &     - \\ \hline
UNetUp (VGG11) & 39.16 & 58.72 & 33.86 & 50.66 & 42.26 \\ \hline
UNetUp (VGG16)  &\textbf{ 41.13} & \textbf{60.04}& 36.12& 53.86&\textbf{44.99} \\ \hline
UNetUp (ResNet18) & 37.48 & 58.08  &33.64 & 52.62 & 41.04\\ \hline
UNetUp (ResNet34)  & 38.97 & 58.39 & 34.69 & 52.94 & 41.91 \\ \hline
\end{tabular}
\end{table*}

\begin{table}[]
\caption{Model Size}
\label{tab:model_size}
\begin{tabular}{|l|l|l|}
\hline
\textbf{Model}                &\textbf{ \# params } & \textbf{Size} \\ \hline
ENetSeparable        & \textbf{226,596 }   & \textbf{1.1MB }       \\ \hline
ENet                 & 349,068    & 1.6MB        \\ \hline
LinkNet (ResNet18)   & 11,686,561 & 46.8MB       \\ \hline
UNetUp (ResNet18) & 20,290,377 & 81.2MB \\ \hline
LinkNet (ResNet34)   & 21,794,721 & 87.3MB       \\ \hline
UNet (ResNet18)   & 22,383,433 & 89.6MB         \\ \hline
UNetUp (VGG11) & 22,927,393 &   91.7MB     \\ \hline
UNet (VGG11)         & 25,364,513 & 101.5MB      \\ \hline
UNetUp (VGG16) & 29,306,465 & 117.2  MB      \\ \hline
UNetUp (ResNet34) & 30,398,537 & 121.7MB \\ \hline
UNet (VGG16)         & 32,202,337 & 128.8MB      \\ \hline
UNet (ResNet34)    & 32,491,593 &  130.1MB\\ \hline
\end{tabular}
\end{table}

It is interesting to note that the model performance was not directly correlated to model size. For instance, in the binary segmentation task in Table \ref{tab:seg_results}, it can be seen that ENetSeparable outperformed ENet even though it had 30\% fewer parameters. The number of parameters in these models were smaller by two orders of magnitude compared to all the other models tested, but for the binary segmentation task, these models were close to the top performing UNetUp (VGG16) model. However, the tradeoff between model size and capacity became obvious in the multiclass segmentation task, where the smaller models did not converge. 

From the results it can be seen that the VGG-based encoders outperformed the ResNet-based encoders for all tasks. For instance, it can be seen from Table \ref{tab:seg_results} that for the road segmentation task using the UNet models, the VGG11 and VGG16 encoders were consistently better than the ResNet18 and ResNet34 encoders. This performance difference can also be seen across the building segmentation and the multi-class segmentation tasks.

The major difference between the UNet models and the LinkNet models is the way the skip link features are treated. In the former, the skip link features are concatenated with the corresponding decoder features whereas in the latter they are added to the decoder features to make the process more efficient. From the results, it can be observed that the feature concatenation in the UNet models allowed the network to learn more discriminative features as these models always outperformed their LinkNet equivalents, even when the encoder was the same. 

\input{building_images.tex}
\input{road_images.tex}

Finally, the proposed UNetUp with a VGG16 encoder outperformed all other models on the segmentation tasks. It could also be seen that the UNetUp models outperformed equivalent UNet models when controlled for the encoder even though they had fewer parameters since they used upsampling instead of transposed convolution layers.

\begin{figure*}[]
\centering
\begin{tabular}{|c|c|c|}
\hline
&  & \\
\begin{subfigure}[b]{0.2\textwidth}
\includegraphics[width = \textwidth]{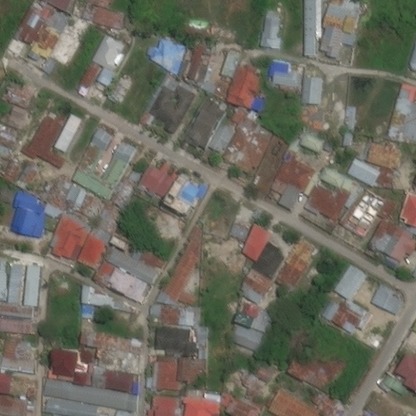} 
\caption{Image}
\end{subfigure} &
\begin{subfigure}[b]{0.2\textwidth}
\includegraphics[width = \textwidth]{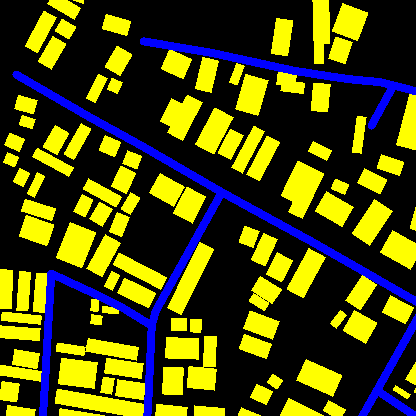} 
\caption{Ground Truth}
\end{subfigure} &
\begin{subfigure}[b]{0.2\textwidth}
\includegraphics[width = \textwidth]{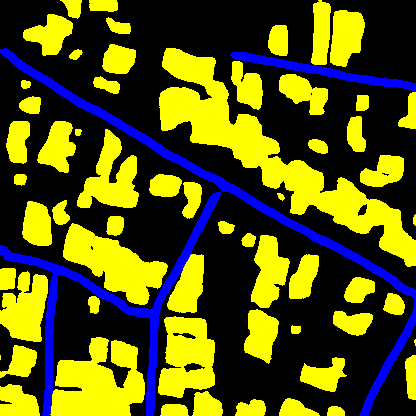} 
\caption{UNetUp (VGG16)}
\end{subfigure} 
\\ \hline
\end{tabular}
\caption{Multi-class segmentation output}
\label{fig:multiclass_seg}
\end{figure*}

Table \ref{tab:seg_results} also shows that all the models had better performance for the binary segmentation task as compared to multi-class segmentation for the same classes. This seems to imply that more training data does not necessarily improve the performance if the task is more complex. An example of the results of the UNetUp (VGG16) model on the multi-class segmentation problem is seen in Figure \ref{fig:multiclass_seg}. The sample image is the same as the one shown for the building segmentation case in Figure \ref{fig:building_seg} and by comparing the two, it can be seen that the results in the multi-class case are less distinctive and more blob like.

\subsection{Quantitative Disaster Mapping Results}

The precision-recall results of the obtained road networks are given in Table \ref{tab:pre_rec}. The road networks created from the segmentation mask of the post-disaster image have been denoted as \textit{Post}. The results of the proposed method where the OSM road network were updated by removing all destroyed road segments are given as \textit{Diff}.

\begin{table*}[]
\caption{Precision Recall of Sub-segments}
\label{tab:pre_rec}
\centering
\begin{tabular}{|l|l|l|l|l|l|l|l|}
\hline
\multicolumn{1}{|c|}{\textbf{Model}} & \multicolumn{2}{c|}{\textbf{Precision}} & \multicolumn{2}{c|}{\textbf{Recall}} & \multicolumn{2}{c|}{\textbf{F\_score}} \\ \hline
\textbf{} & \textbf{Post} & \textbf{Diff} & \textbf{Post} & \textbf{Diff} & \textbf{Post} & \textbf{Diff} \\ \hline
ENet & 70.36 & 92.49 & 65.97 & 93.13 & 68.09 & 92.81 \\ \hline
ENetSeparable & 61.20 & 92.65 & 67.12 &\textbf{ 95.5} & 64.02 & 94.05 \\ \hline
LinkNet (ResNet18) & 77.02 & 93.8 & 70.36 & 95.29 & 73.54 & 94.54 \\ \hline
LinkNet (ResNet34) & 71.8 & \textbf{95.18} & \textbf{76.29 }& 94.34 & \textbf{73.98} & \textbf{94.76} \\ \hline
UNet (VGG11) & 73.66 & 94.8 & 67.87 & 92.8 & 70.65 & 93.8 \\ \hline
UNet (VGG16) & 75.73 & 94.99 & 72.3 & 94.02 & \textbf{73.98 }& 94.5 \\ \hline
UNet (ResNet18) &\textbf{ 81.71} & 93.64 & 63.86 & 92.27 & 71.69 & 92.95 \\ \hline
UNet (ResNet34) & 78.32 & 94.09 & 74.57 & 95.41 & 76.40 & 94.75  \\ \hline
UNetUp (VGG11) & 71.30 & 94.4 & 67.31 & 93.93 & 69.25 & 94.16 \\ \hline
UNetUp (VGG16) & 67.17 & 95.1 & 70.93 & 93.68 & 69.0 & 94.38 \\ \hline
UNetUp (ResNet18) & 78.17 & 94.32 & 67.42 & 95.12 & 72.40 & 94.72 \\ \hline
UNetUp (ResNet34) & 76.4 & 93.76 & 67.14 & 93.67 & 71.47 & 93.71 \\ \hline
\end{tabular}
\end{table*}

The \textit{Post} results convey the generalisation capability of the tested networks across image datasets from different times since they were trained on pre-disaster imagery, while the evaluation was over the post-disaster imagery. In contrast to the pre-disaster results, the best performing model for precision-recall was UNet with a ResNet backend.

It can be seen that in the case of \textit{Post}, the precision was usually much higher than the recall implying that the segmentation network has a higher number of false negatives than false positives. This was due to the fact that there were gaps in the segmentation mask caused due to occlusions from shadows, buildings, etc. LinkNet with a ResNet34 backend gave the highest recall in this case.

As Table \ref{tab:pre_rec} shows, the proposed \textit{Diff} framework helped improve the generated road graph, regardless of the base network used. The difference in results between the various architectures also became less pronounced as can be seen in Table \ref{tab:pre_rec} where the difference between the maximum and minimum F\_score in the case of \textit{Post} was approximately 8\% whereas that for \textit{Diff} was 2\%. This was largely due to the fact that the proposed method benefited from prior knowledge from OSM. Note that the OSM data is not completely accurate~\cite{Mattyus2017}. However, based on empirical observations, using this data provides significantly better results than assuming no prior knowledge.


\begin{table*}[]
\caption{Connectivity results. All numbers given as percentage.}
\centering
\label{tab:conn_res}
\begin{tabular}{|l|l|l|l|l|l|l|l|l|l|}
\hline
\textbf{Model} & \multicolumn{2}{c|}{\textbf{Correct}} & \multicolumn{2}{c|}{\textbf{Too Long}} & \multicolumn{2}{c|}{\textbf{Too Short}} & \multicolumn{2}{c|}{\textbf{No Connection}} \\ \hline
\textbf{} & \textbf{Post} & \textbf{Diff} & \textbf{Post} & \textbf{Diff} & \textbf{Post} & \textbf{Diff} & \textbf{Post} & \textbf{Diff} \\ \hline
ENet & 21.82 & 54.02 & 19.15 & 14.81 & 2.29 & 1.26 & 56.66 & 29.83 \\ \hline
ENetSeparable & 20.29 & 62.22 & 18.43 & 14.94 & 2.18 & 1.35 & 59.04 & 21.39 \\ \hline
LinkNet (ResNet18) & 18.33 & 70.36 & 23.65 & 6.93 & 2.35 & 1.35 & 55.59 & 21.25 \\ \hline
LinkNet (ResNet34) & 31.05 & 74.59 & 14.97 & 5.2 & 3.97 & 1.41 & 49.96 & 18.71 \\ \hline
UNet (VGG11) & 25.15 & 65.56 & 25.58 & 9.99 & 5.28 & 1.02 & 43.94 & 23.33 \\ \hline
UNet (VGG16) & 26.8 & 67.68 & 31.64 & 8.09 & 5.42 & \textbf{0.97} & \textbf{36.4 }& 23.16 \\ \hline
UNet (ResNet18) & 21.03 & 58.62 & 20.15 & 12.57 & 4.72 & 1.07 & 54 & 27.65 \\ \hline
UNet (ResNet34) &\textbf{ 37.04 }& \textbf{75.58 }& 17.13 & \textbf{4.41 }& 4.68 & 1.43 & 41.1 & \textbf{18.49} \\ \hline
UNetUp (VGG11) & 19.58 & 61.03 & \textbf{14.22} & 13.58 & \textbf{1.94 }& 0.99 & 64.23 & 24.3 \\ \hline
UNetUp (VGG16) & 24.3 & 66.18 & 23.5 & 8.98 & 4.71 & 1.09 & 47.43 & 23.66 \\ \hline
UNetUp (ResNet18) & 22.44 & 67.35 & 21.64 & 8.46 & 2.66 & 1.14 & 53.23 & 23.05 \\ \hline
UNetUp (ResNet34) & 18.71 & 66.01 & 23.9 & 8.89 & 2.24 & 1.57 & 55.1 & 23.44 \\ \hline
\end{tabular}
\end{table*}






The connectivity results of the estimated post-disaster road networks are reported in Table \ref{tab:conn_res}. Similar to the precision-recall results, it can be seen that the proposed framework improved the results by a large margin. This was due to the fact that the output of the segmentation networks often had gaps which caused missing connections in the generated road networks. The use of the OSM network, which is properly connected, as an initial estimate, helped deal with these missing connections. This conjecture is supported by the number of pairs that are marked as having 'No Connections' in Table \ref{tab:conn_res} where using the \textit{Diff} framework reduced the number of 'No Connection' pairs to half of those from \textit{Post}. The \textit{Post} results had a number of small disconnected segments and some spurious paths caused due to a non-ideal segmentation mask. The \textit{Diff} results, on the other hand, were much better connected. However, \textit{Diff} did have some incorrect segments where the mask difference missed segments.



\subsection{Qualitative Disaster Impact Results}

\begin{figure}[]
    \centering
    \includegraphics[width=0.5\textwidth]{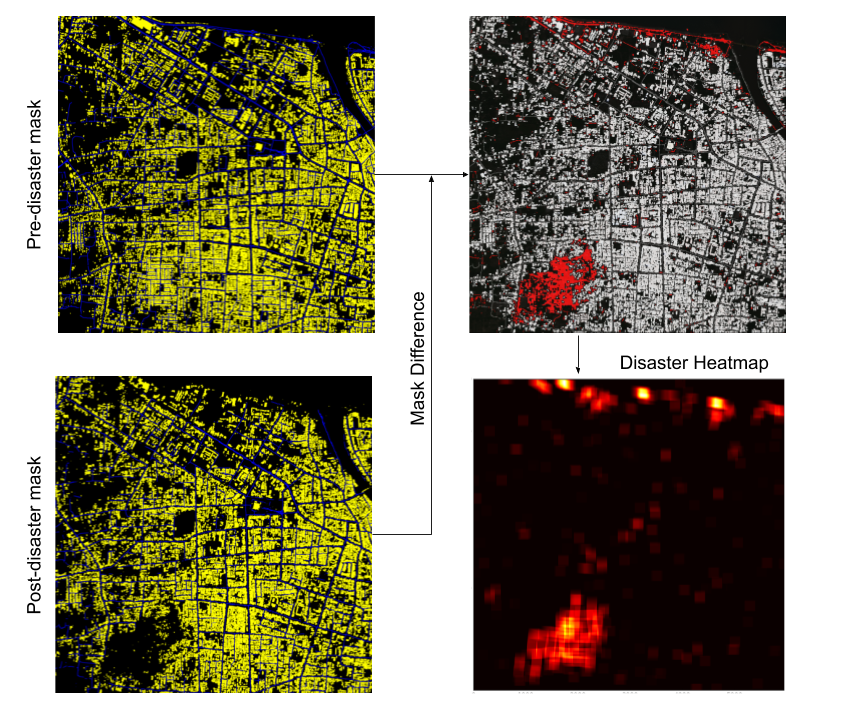}
    \caption{Disaster Impact assessment. \textit{Left}: Road and building masks from satellite imagery using segmentation network with buildings in yellow and roads in blue; \textit{Top Right}: Estimated difference between the infrastructure before and after disaster given in red; \textit{Bottom Right}: Change heatmap overlaid onto an image of the test region. }
    \label{fig:disaster_impact}
\end{figure}

As outlined in Section \ref{sec:mask_diff}, the difference between the segmentation masks from  pre-disaster and post-disaster imagery can be used for disaster impact assessment. This process is shown in Figure \ref{fig:disaster_impact} where the difference in the buildings and roads caused by the disaster are marked in red in the image on the top right.  

The area under consideration was divided into a grid of cells of a fixed size and the number of changed pixels per grid cell was used as an overall estimate of the damage caused to a particular area. This has been plotted as a heatmap in Figure \ref{fig:disaster_impact}. The heatmap shows that the major destruction was along the coast and an area in the south-west of Palu city. This finding corresponds to the European Commission's (EC) Corpernicus Emergency Mapping Services report\cite{Centre2018} for the area. A major portion of the coast was washed away due to the tsunami and the south-west region of the city was washed away due to soil liquefaction.

{}

\section{Conclusions}

This work provides a comparison among different segmentation models and presents a framework for the identification of damaged areas and accessible roads in post-disaster scenarios using satellite imagery. The framework leverages on pre-existing knowledge from OSM to obtain a robust estimate of the affected road network.

The performances of various models for the tasks of binary and multi-class semantic segmentation in aerial images have been analysed and compared. The results show that using encoders pretrained on ImageNet improved the training time by around 10 epochs and the accuracy by a couple of percentage points despite the domain gap that existed between ImageNet and aerial images.

On comparing the effects of using different encoders for the task of semantic segmentation, it could be seen that VGG16 outperformed all other feature extraction modules. The tradeoff between accuracy and efficiency has been studied. An extremely efficient neural network, termed ENetSeparable, was proposed. It has  30\% fewer parameters than ENet and still performed better on the binary segmentation task.

For post-disaster scenarios, areas affected by the disaster were identified using the difference in the predicted segmentation masks. The evaluated road changes were used to update the road networks available from OSM. There was a significant difference in the results of the various segmentation networks where the F\_score varied by as much as 8\%. The use of the proposed framework alleviated the differences and brought the difference in F\_score down to 2\%. The highest F\_score achieved with the use of the proposed framework was 94.76 as compared to the highest F\_score of 73.98 from the segmentation networks. 

The proposed framework uses OSM data for training and does not require time-consuming manually annotated post-disaster data. Finally, the qualitative assessment of the aftermath damage can be generated easily, as shown in the Palu tsunami, which was validated from the European Commission report. 

This work can be further improved in a number of ways. Namely, the results of the different models could be ensembled to help improve the road connectivity results. Classification of damages could be used to identify where the infrastructure has been completely destroyed, as in the case of soil liquefaction, or if a road blockage is something that can be dealt with relatively easily, such as one caused by a fallen tree.





\section*{ACKNOWLEDGMENT}

The authors would like to thank Dr. Andrew West, Dr. Thomas Wright and Ms. Elisabeth Welburn for their valuable comments and feedback. A. Gupta is funded by a Scholarship from the Department of Electrical and Electronic Engineering, The University of Manchester and the ACM SIGHPC/Intel Computational and Data Science Fellowship.

\bibliographystyle{splncs04}
\bibliography{main}


\end{document}

%% file: building_images.tex
\begin{figure*}[]
\begin{tabular}{|c|c|c|c|}
\cline{0-1}
&  \\
\begin{subfigure}[b]{0.2\textwidth}
\includegraphics[width = \textwidth]{images/buildings/0_RGB-PanSharpen_3230213-pre-86.jpg} 
\caption{Image}
\end{subfigure} &
\begin{subfigure}[b]{0.2\textwidth}
\includegraphics[width = \textwidth]{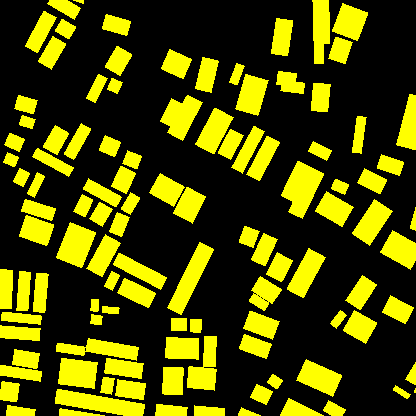} 
\caption{Ground Truth} 
\end{subfigure} 
\\ \hline

& & & \\

\begin{subfigure}[b]{0.2\textwidth}
\includegraphics[width = \textwidth]{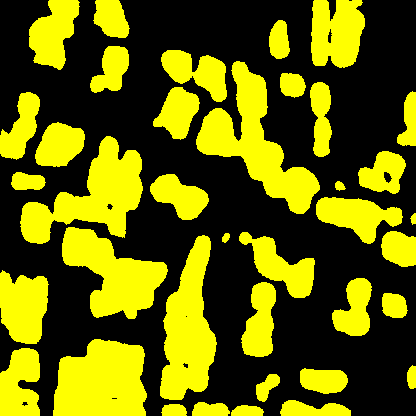} 
\caption{ENet}
\end{subfigure} &
\begin{subfigure}[b]{0.2\textwidth}
\includegraphics[width = \textwidth]{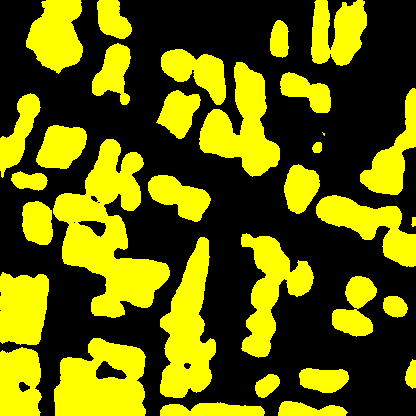}
\caption{ENetSeparable}
\end{subfigure}  &
\begin{subfigure}[b]{0.2\textwidth}
\includegraphics[width = \textwidth]{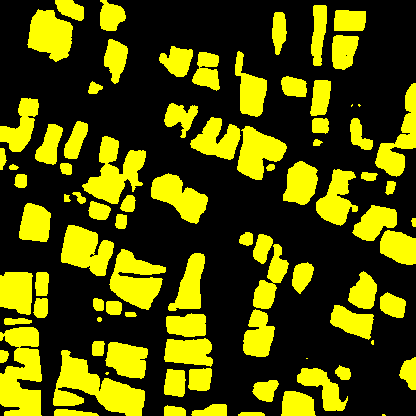}
\caption{LinkNet (ResNet18)}
\end{subfigure} &  
\begin{subfigure}[b]{0.2\textwidth}
\includegraphics[width = \textwidth]{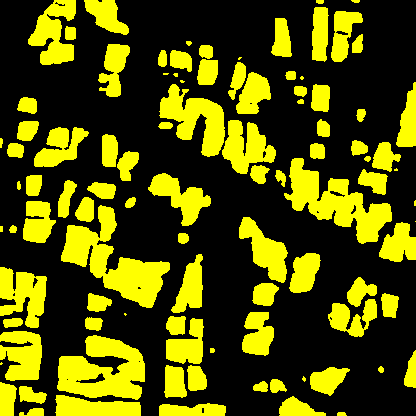}
\caption{LinkNet (ResNet34)}
\end{subfigure}  \\ \hline 
& & & \\

\begin{subfigure}[b]{0.2\textwidth}
\includegraphics[width = \textwidth]{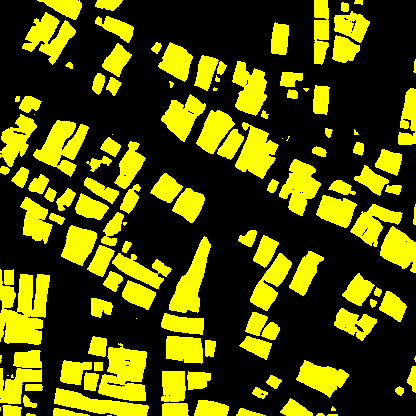}
\caption{UNet (VGG11)}
\end{subfigure} &
\begin{subfigure}[b]{0.2\textwidth}
\includegraphics[width = \textwidth]{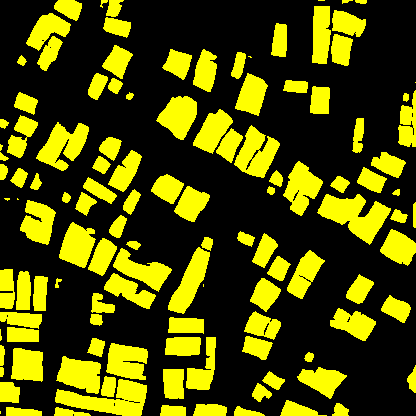}
\caption{UNet (VGG16)}
\end{subfigure}  &
\begin{subfigure}[b]{0.2\textwidth}
\includegraphics[width = \textwidth]{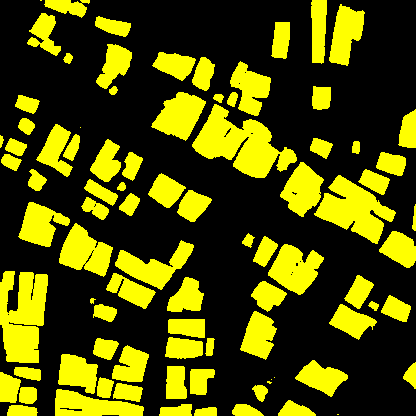}
\caption{UNet (ResNet18)}
\end{subfigure}  &
\begin{subfigure}[b]{0.2\textwidth}
\includegraphics[width = \textwidth]{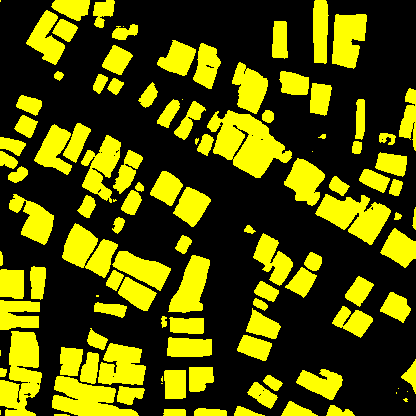}
\caption{UNet (ResNet34)}
\end{subfigure} 
\\ \hline & & & \\

\begin{subfigure}[b]{0.2\textwidth}
\includegraphics[width = \textwidth]{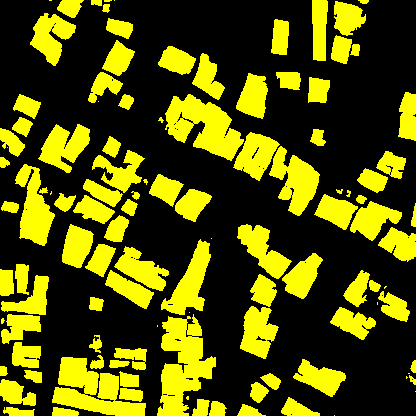}
\caption{UNetUp (VGG11)}
\end{subfigure} &
\begin{subfigure}[b]{0.2\textwidth}
\includegraphics[width = \textwidth]{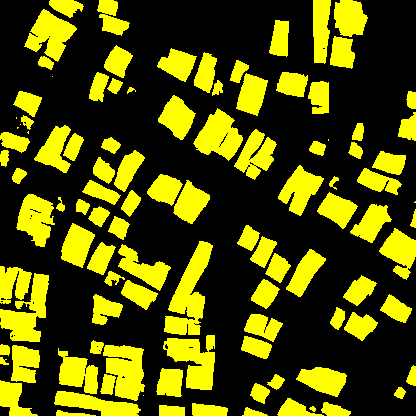}
\caption{UNetUp (VGG16)}
\end{subfigure} &
\begin{subfigure}[b]{0.2\textwidth}
\includegraphics[width = \textwidth]{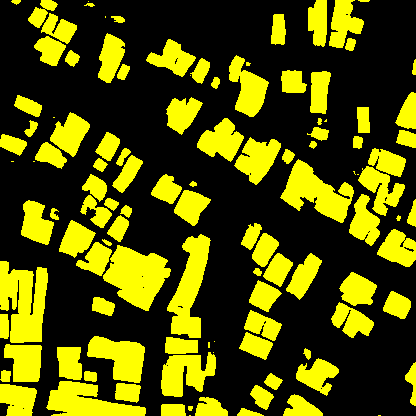}
\caption{UNetUp (ResNet18)}
\end{subfigure}&
\begin{subfigure}[b]{0.2\textwidth}
\includegraphics[width = \textwidth]{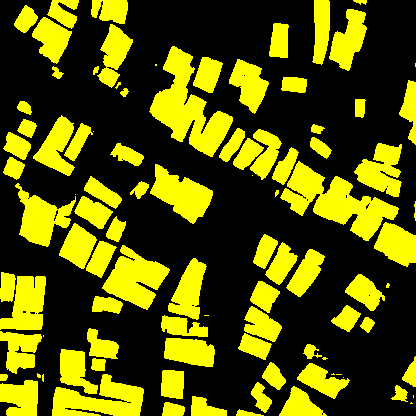}
\caption{UNetUp (ResNet34)}
\end{subfigure} \\ \hline

\end{tabular}
\caption{Visualisation of the building segmentation results using pretrained encoders}
\label{fig:building_seg}
\end{figure*}

%% file: road_images.tex
\begin{figure*}[]
\begin{tabular}{|c|c|c|c|}
\cline{0-1}
&  \\
\begin{subfigure}[b]{0.2\textwidth}
\includegraphics[width = \textwidth]{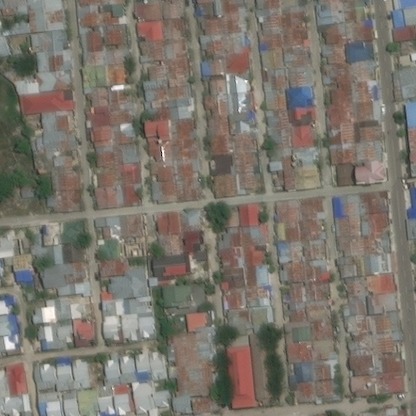} 
\caption{Image}
\end{subfigure} &
\begin{subfigure}[b]{0.2\textwidth}
\includegraphics[width = \textwidth]{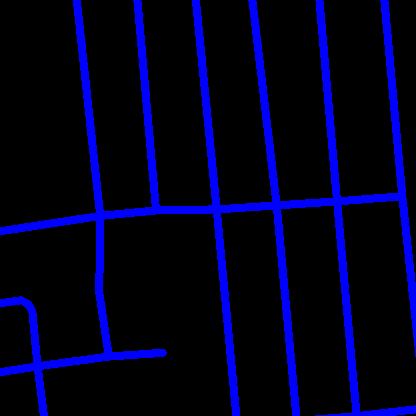} 
\caption{Ground Truth} 
\end{subfigure} 
\\ \hline
& & & \\

\begin{subfigure}[b]{0.2\textwidth}
\includegraphics[width = \textwidth]{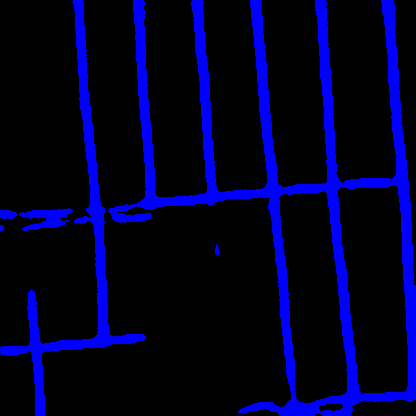} 
\caption{ENet}
\end{subfigure} &
\begin{subfigure}[b]{0.2\textwidth}
\includegraphics[width = \textwidth]{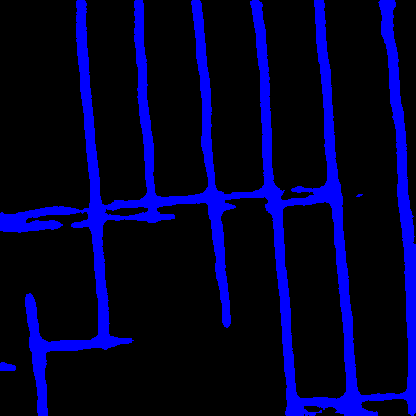}
\caption{ENetSeparable}
\end{subfigure} &
\begin{subfigure}[b]{0.2\textwidth}
\includegraphics[width = \textwidth]{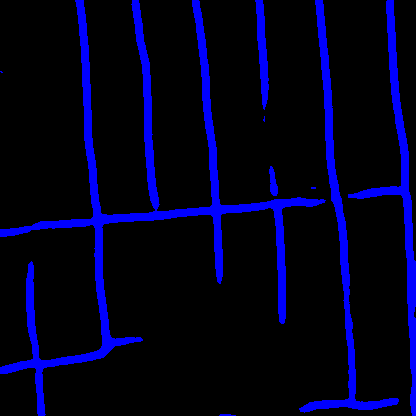}
\caption{LinkNet (ResNet18)}
\end{subfigure} &  
\begin{subfigure}[b]{0.2\textwidth}
\includegraphics[width = \textwidth]{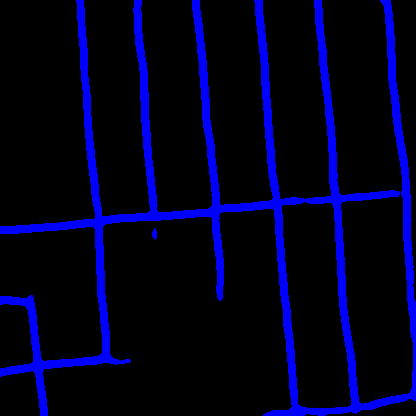}
\caption{LinkNet (ResNet34)}
\end{subfigure}  \\ \hline 
& & & \\

\begin{subfigure}[b]{0.2\textwidth}
\includegraphics[width = \textwidth]{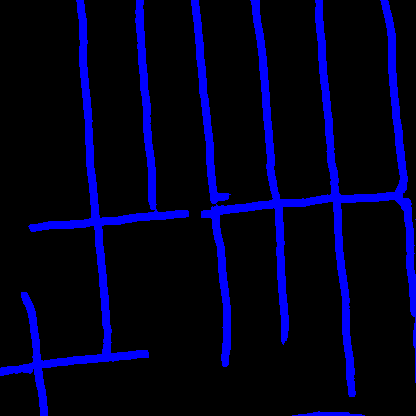}
\caption{UNet (VGG11)}
\end{subfigure} &
\begin{subfigure}[b]{0.2\textwidth}
\includegraphics[width = \textwidth]{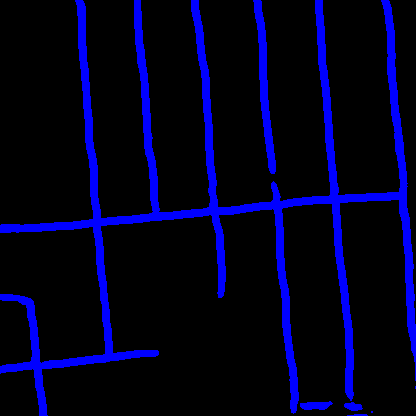}
\caption{UNet (VGG16)}
\end{subfigure} &
\begin{subfigure}[b]{0.2\textwidth}
\includegraphics[width = \textwidth]{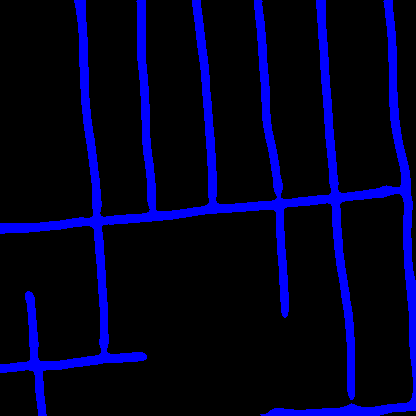}
\caption{UNet (ResNet18)}
\end{subfigure} &
\begin{subfigure}[b]{0.2\textwidth}
\includegraphics[width = \textwidth]{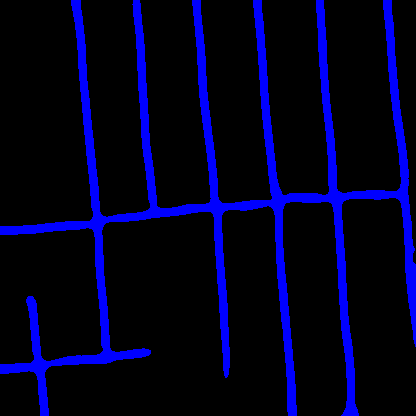}
\caption{UNet (ResNet34)}
\end{subfigure} \\ \hline & & & \\

\begin{subfigure}[b]{0.2\textwidth}
\includegraphics[width = \textwidth]{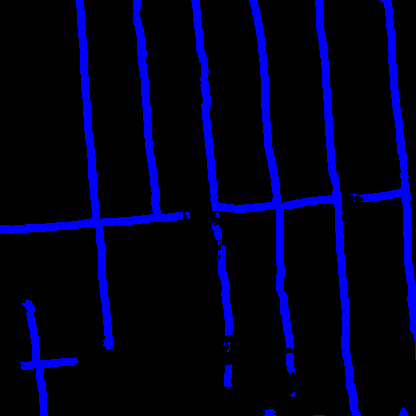}
\caption{UNetUp (VGG11)}
\end{subfigure} &
\begin{subfigure}[b]{0.2\textwidth}
\includegraphics[width = \textwidth]{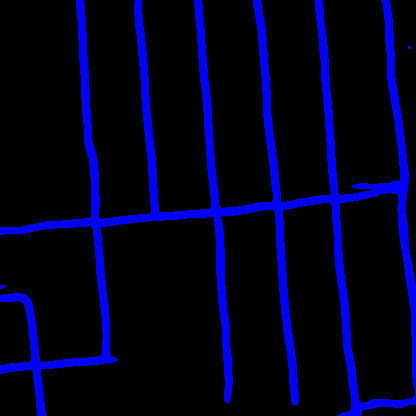}
\caption{UNetUp (VGG16)}
\end{subfigure}  &
\begin{subfigure}[b]{0.2\textwidth}
\includegraphics[width = \textwidth]{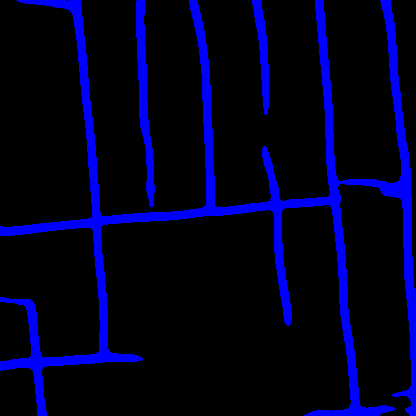}
\caption{UNetUp (ResNet18)}
\end{subfigure} &
\begin{subfigure}[b]{0.2\textwidth}
\includegraphics[width = \textwidth]{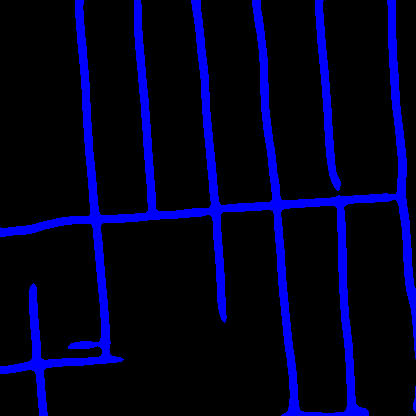}
\caption{UNetUp (ResNet34)}
\end{subfigure} 
\\ \hline

\end{tabular}
\caption{Visualisation of the road segmentation results using pretrained encoders}
\label{fig:road_seg}
\end{figure*}